\documentclass[preprint,showpacs,preprintnumbers,amsmath,amssymb]{revtex4}
\usepackage{epsf}
\usepackage{dcolumn}% Align table columns on decimal point
\usepackage{bm}% bold math
\everymath{\displaystyle}
\begin{document}

\title{Analysis of wave equations for spin-1 particles interacting with an
electromagnetic field}

\author{Alexander J. Silenko}

\affiliation{Institute of Nuclear Problems, Belarusian State
University, 220080 Minsk, Belarus}

\date{\today}

\begin {abstract}
The Foldy-Wouthuysen transformation for relativistic spin-1
particles interacting with nonuniform electric and uniform
magnetic fields is performed. The Hamilton operator in the
Foldy-Wouthuysen representation is determined. It agrees with the
Lagrangian obtained by Pomeransky, Khriplovich, and Sen'kov. The
classical and quantum formulae for the Hamiltonian agree. The
validity of the Corben-Schwinger equations is confirmed. However,
it is difficult to generalize these equations in order to take
into account the quadrupole moment defined by a particle charge
distribution. The known second-order wave equations are not quite
satisfactory because they contain non-Hermitian terms. The
Hermitian second-order wave equation is derived.
\end{abstract}

\maketitle

\section {Introduction}

   The investigation of interaction of spin-1 particles with an electromagnetic
field is very important for the high energy physics. There exist
some difficulties in the spin-1 particle physics. There are many
works where a consistency of spin-1 particle theories has been
considered (see Ref. \cite{VSM} and references therein). However,
these works have not given us final conclusions. The method
elaborated by Pomeransky, Khriplovich, and Sen'kov has made it
possible to describe spin effects in an interaction of particles
of any spin with an electromagnetic field \cite{PK,PS}. In
particular, the Lagrangian with an allowance for second-order
terms in spin has been calculated \cite{PK,PS}. In the present
work, we use these results to verify some wave equations for
spin-1 particles. We transform the Hamilton operator to the
block-diagonal form (diagonal in two spinors) which defines the
Foldy-Wouthuysen (FW) representation \cite{FW}. This
representation is very convenient in order to analyze spin effects
and perform the semiclassical transition. The obtained result is
compared with both classical \cite{Ycl} and
Pomeransky-Khriplovich-Sen'kov (PKS) \cite{PK,PS} approaches.

\section {Equations for spin-1 particles}

   The situation in the spin-1 particle theory differs essentially from the
situation in the spin-0 and spin-1/2 particle theories. The
important difference is a great number of equations describing
spin-1 particles (vector mesons). For the first time, the
equations for vector mesons have been found by Proca \cite{Pr}.
For particles in an electromagnetic field, they have the form:
\begin{equation} U_{\mu\nu}=D_\mu U_\nu-D_\nu U_\mu, ~~~
D^\mu U_{\mu\nu}=m^2 U_\nu, ~~~ D_\mu=\partial_\mu+ieA_\mu\equiv
\left(\frac {\partial}{\partial t}+ie\Phi,\nabla-ie\bm A\right),
\label{eq1} \end{equation} where $A_\mu,\Phi$, and $\bm A$ are
4-potential, scalar potential, and vector potential of
electromagnetic field, respectively.

Spin-1 particles can be also described by the
Duffin-Kemmer-Petiaux (DKP) \cite{D,K,Pe}, Stuckelberg \cite{St},
multispinor Bargmann-Wigner \cite{BW}, and other equations. The
DKP equation has the form
$$ (\beta_\mu D_\mu+m)\psi=0. $$

In this equation, the wave function $\psi$ has ten components, and
$\beta_ \mu$ are 10$\times$10 matrices. They satisfy the
conditions:
$$ \beta_\mu \beta_\nu \beta_\lambda+\beta_\lambda \beta_\nu \beta_\mu=
\beta_\mu \delta_{\nu\lambda}+\beta_\lambda\delta_{\mu\nu}. $$

Corben and Schwinger \cite{CS} showed how to include an anomalous
magnetic dipole term in the Proca equations. Young and Bludman
\cite{YB} took into account the additional electric quadrupole moment
defined by a charge distribution in particles (charge quadrupole moment).

Many first-order equations are equivalent. They can be transformed
one into another. This refers to the Proca, DKP, and
Bargmann-Wigner equations (see Refs. \cite{YB,Um}). The
Stuckelberg equations \cite{St} differ essentially from other
equations by the inclusion of an additional scalar field. As a
result, the corresponding wave function has eleven components. The
wave functions of the Proca and DKP equations have ten components.

Kahler \cite{Kah} has proposed the equation for inhomogeneous
differential forms which is equivalent to a system of scalar,
vector, antisymmetric tensor, pseudovector and pseudoscalar fields
\cite{Kr1,Kr2}. Kruglov \cite{Kr3} has generalized this equation
on the case when the mass of scalar and pseudoscalar fields,
$m_0$, differs from the mass of other fields, $m$. The
corresponding wave function has sixteen components.

Several components of the Proca equations can be expressed in
terms of the others. As a result, the equations for the
ten-component wave function can be reduced to the equation for the
six-component one (the generalized Sakata-Taketani equation
\cite{YB,SaTa}). Since the components of the reduced wave function
are two spinors, the wave function of the generalized
Sakata-Taketani equation is a bispinor. This equation is very
convenient for the semiclassical transition simplifying an
investigation of spin dynamics.

   Besides the first-order wave equations, there exist also second-order ones.
   These are the second-order forms of the above
mentioned equations and some other equations (e.g., the Shay-Good
equation \cite{SG}). The Shay-Good equation is not equivalent to
the Proca theory.

\section {Consistency of spin-1 particle theories}

  Soon after the appearance of the Proca theory, the problem of its consistency
was stated \cite{Ta}. There are many works where several
difficulties of spin-1 particle theories have been investigated
(e.g., complex energy modes for particles in a uniform magnetic
field, see Refs. \cite{VSM,TY,GT,T,Ts} and references therein). In
these works the problem of consistency of spin-1 particle theories
was solved qualitatively. However, there exists a more exact
criterium of validity of any particle theory. As is shown in Refs.
\cite{PK,PS,Zw}, the spin motion of particles with arbitrary spin
is described by the Bargmann-Michel-Telegdi (BMT) equation
\cite{BMT}:
\begin{equation}  \begin{array}{c}
\left(\frac{d\bm S}{dt}\right)_{BMT}=\frac{e}{2m}
\left\{\left(g-2+\frac{2}{\gamma}\right)[\bm S\times\bm B]-(g-2)
\frac{\gamma}{\gamma+1}[\bm S\times\bm v](\bm v\cdot \bm B)
    \right. \\ \left.
+\left(g-2+\frac{2}{\gamma+1}\right)\left[\bm S\times [\bm
E\times\bm v] \right]\right\},
\end{array} \label{eq2} \end{equation}
where $\bm S$ is the spin operator, $\bm E$ is the electric field
strength, and $\bm B$ is the magnetic field induction. The
equation for the unit polarization vector, $\bm O=<\bm S>/S$, has
the same form.

Any wave equation should be in congruence with the BMT equation.

To verify wave equations for spin-1 particles, the Lagrangian
obtained in Refs. \cite{PK,PS} will also be used. This Lagrangian
describes spin effects for particles of an arbitrary spin
interacting with an electromagnetic field. It is given by
\begin{equation} \begin{array}{c}
{\cal L}={\cal L}_1+{\cal L}_2,\\
{\cal
L}_1=\frac{e}{2m}\left\{\left(g-2+\frac{2}{\gamma}\right)(\bm
S\cdot \bm B)-(g-2)\frac{\gamma}{\gamma+1}(\bm S\cdot\bm v)(\bm
v\cdot\bm B) \right. \\ \left.
+\left(g-2+\frac{2}{\gamma+1}\right)(\bm S\cdot[\bm E\times\bm
v])\right\},
\\ {\cal L}_2=\frac{Q}{2S(2S-1)}\left[(\bm S\cdot\nabla)-
\frac{\gamma}{\gamma+1}(\bm S\cdot\bm v)(\bm v\cdot\nabla)\right]
\left[(\bm S\cdot\bm E)-\frac{\gamma}{\gamma+1}(\bm S\cdot\bm v)
(\bm v\cdot\bm E) \right. \\
+(\bm S\cdot[\bm v\times\bm B])\Biggr]+
\frac{e}{2m^2}\frac{\gamma}{\gamma+1}(\bm S\cdot[\bm
v\times\nabla])\left[ \left(g-1+\frac{1}{\gamma}\right)(\bm
S\cdot\bm B) \right. \\ \left. -(g-1)\frac{\gamma}{\gamma+1}(\bm
S\cdot\bm v)(\bm v\cdot\bm B)+
\left(g-\frac{\gamma}{\gamma+1}\right)(\bm S\cdot[\bm E\times\bm
v])\right],\\ \gamma=\frac{1}{\sqrt{1-\bm
v^2}}=\frac{\sqrt{m^2+\bm\pi^2}}{m},
\end{array} \label{eq3} \end{equation}
where $g=2\mu m/(eS)$, $Q$ is the quadrupole moment and $\gamma$
is the Lorentz factor. In Lagrangian (3), ${\cal L}_1$ contains
terms that are linear in spin, while ${\cal L}_2$ contains
quadratic terms. The Hermitian form of relation (3) is obtained by
the substitution $${\cal L}\rightarrow ({\cal L}+{\cal
L}^{\dag})/2.$$

The corresponding Hamiltonian equals this Lagrangian with the
opposite sign: ${\cal H}=-{\cal L}$. In Eq. (3), the velocity
operator is replaced by the corresponding classical quantity, and
the spin is described by appropriate spin matrices. Therefore,
Lagrangian (3) has been obtained in the semiclassical
approximation. This Lagrangian has been used for finding the
general equation of spin motion in nonuniform fields \cite{YP}:
\begin{equation}
\frac{d\bm S}{dt}=\left(\frac{d\bm S}{dt}\right)_{BMT}+
\left(\frac{d\bm S}{dt}\right)_{q}, \label{eq4} \end{equation}
\begin{equation} \begin{array}{c}
\left(\frac{d\bm S}{dt}\right)_{q}=
\frac{Q}{4S(2S-1)}\left(\Biggl\{\biggl([\bm S\times\nabla]-
\frac{\gamma}{\gamma+1}[\bm S\times\bm v](\bm
v\cdot\nabla)\biggr), \biggl((\bm S\cdot\bm E)-
 \right. \\ \left. \frac{\gamma}{\gamma+1}(\bm S\cdot\bm v)
(\bm v\cdot\bm E)+ (\bm S\cdot[\bm v\times\bm
B])\biggr)\Biggr\}_++\Biggl\{ \biggl((\bm S\cdot\nabla)-
\frac{\gamma}{\gamma+1}(\bm S\cdot\bm v)(\bm
v\cdot\nabla)\biggr), \biggl([\bm S\times\bm E] \right. \\
\left. -\frac{\gamma}{\gamma+1}[\bm S\times\bm v] (\bm v\cdot\bm
E)+\left[
\bm S\times[\bm v\times\bm B]\right]\biggr)\Biggr\}_+\right) \\
+\frac{e}{4m^2}\frac{\gamma}{\gamma+1}\left(\left\{\left[\bm
S\times [\bm v\times\nabla]\right],\left[
\left(g-1+\frac{1}{\gamma}\right)(\bm S\cdot \bm
B)-(g-1)\frac{\gamma}{\gamma+1}(\bm S\cdot\bm v)(\bm v\cdot\bm B)
\right.\right.\right. \\ \left.\left.\left.
+\left(g-\frac{\gamma}{\gamma+1}\right)(\bm S\cdot[\bm E\times\bm
v]) \right]\right\}_++ \left\{\biggl(\bm S\cdot [\bm
v\times\nabla]\biggr),\left[
\biggl(g-1+\frac{1}{\gamma}\biggr)[\bm S\times\bm B]
\right.\right.\right. \\ \left.\left.\left.
-(g-1)\frac{\gamma}{\gamma+1}[\bm S\times\bm v](\bm v\cdot\bm B)+
\biggl(g-\frac{\gamma}{\gamma+1}\biggr)\left[\bm S\times[\bm
E\times\bm v] \right]\right]\right\}_+\right),
\end{array} \label{eq5} \end{equation}
where $\{\dots,\dots\}_+$ means an anticommutator.

The average product $<S_iS_j>$ is not equal to $<S_i><S_j>$. The
quantities $\left(\frac{d\bm S}{dt}\right)_{BMT}$ and
$\left(\frac{d\bm S}{dt}\right)_{q}$ characterize the spin motion
determined by the terms linear [BMT equation (2)] and
quadratic [Eq. (5)] in spin, respectively. Lagrangian (3) and Eqs.
(2),(4),(5) characterize the semiclassical approximation for the
multispinor theory.

To verify any wave equation, it is helpful to transform it to the
Hamilton form and then fulfil the semiclassical transition. The
usual method of performing such a transition is the FW
transformation \cite{FW}. The comparison of obtained semiclassical
expressions with Eqs. (2)--(5) ensures a good possibility of the
verification.

In the FW representation, the Hamiltonian and all the operators
are block-diagonal (diagonal in two spinors). The relations
between the operators are similar to those between the respective
classical quantities. In this representation, the operators have
the same form as in the nonrelativistic quantum theory. Only the
FW representation possesses these properties considerably
simplifying the transition to the semiclassical description. The
FW representation provides the best possibility of obtaining a
meaningful classical limit of the relativistic quantum theory
\cite{CMcK}.

  In the FW representation, the polarization
operator has the simplest form
$$\bm\Pi=\left(\begin{array}{cc} \bm{S}  &  0 \\ 0 &
-\bm{S}\end{array}\right), $$ where $\bm{S}$ is the 3$\times$3
spin matrix for spin-1 particles. In other representations this
operator is expressed by much more cumbersome formulae. Therefore,
other representations are much less convenient in order to find
spin motion equations. This conclusion is valid for particles of
any spin.

For spin-1/2 particles, the polarization operator also takes the
simplest form in the FW representation and cumbersome forms in
other representations. The explicit expressions for this operator
in the Dirac and FW representations are given in Refs.
\cite{FG,JMP}.

   The operator equation of spin motion is determined by the commutator
\begin{equation} \frac{d\bm\Pi}{dt}=i[{\cal H},\bm\Pi].
\label{eq6} \end{equation} To find the equation for the average
spin, Eq. (6) should be averaged.

\section {Foldy-Wouthuysen transformation for spin-1 particles}

   The FW transformation for spin-1 particles has some peculiarities. The wave functions are pseudo-orthogonal,
e.g., their normalization is defined by the relation
$$\int {\Psi^\dag\rho_3\Psi dV}=\int {(\phi^\dag\phi-\chi^\dag\chi)
 dV}=1, $$
where $\Psi =\left(\begin{array}{c} \phi \\ \chi
\end{array}\right)$ is the six-component wave function
(bispinor). Here and below $\rho_i~(i=1,2,3)$ are the Pauli
matrices:
$$\rho_1=\left(\begin{array}{cc}0&1\\1&0\end{array}\right),~~~
\rho_2=\left(\begin{array}{cc}0&-i\\i&0\end{array}\right),~~~
\rho_3=\left(\begin{array}{cc}1&0\\0&-1\end{array}\right).$$
Components of these matrices act on the upper and lower spinors.
The Hamiltonian for spin-1 particles is pseudo-Hermitian, that is,
it satisfies the conditions
$$ {\cal H}=\rho_3{\cal H}^\dag\rho_3, ~~~{\cal H}^\dag=\rho_3{\cal H}\rho_3.
$$
Even (diagonal) terms of the Hamiltonian are Hermitian and odd (off-diagonal) terms are
anti-Hermitian.

The operator $U$ transforming the wave function to a different
representation should be pseudo-unitary:
$$ U^{-1}=\rho_3 U^\dag\rho_3, ~~~U^\dag=\rho_3 U^{-1}\rho_3. $$

   The transformed Hamiltonian equals \cite{FW}:
$$ {\cal H}'=U\biggl({\cal H}-i\frac {\partial}{\partial t}\biggr) U^{-1}+
i\frac {\partial}{\partial t}. $$

The initial Hamiltonian is determined by the generalized
Sakata-Taketani equation which can be written in the form
\begin{equation} {\cal H}=\rho_3 {\cal M}+{\cal E}+{\cal
O},~~~\rho_3 {\cal E}={\cal E}\rho_3, ~~~\rho_3 {\cal O}=-{\cal O}\rho_3,
\label{eq7} \end{equation}
where ${\cal E}$ and ${\cal O}$ are the even and odd operators,
commuting and anticommuting with $\rho_3$, respectively.

When
\begin{equation}
[{\cal M},{\cal O}]=0,~~~ [{\cal E},{\cal O}]=0, \label{eq8}
\end{equation} and the external field is stationary, the exact transformation of
Hamiltonian ${\cal H}$ to the block-diagonal form can be fulfilled
with the operator
\begin{equation}
U=\frac{\epsilon+{\cal M}+\rho_3{\cal
O}}{\sqrt{2\epsilon(\epsilon+{\cal M})}},~~~
U^{-1}=\frac{\epsilon+{\cal M}-\rho_3{\cal
O}}{\sqrt{2\epsilon(\epsilon+{\cal M})}}, ~~~\epsilon=\sqrt{{\cal
M}^2+{\cal O}^2}. \label{eq9} \end{equation} For spin-1/2
particles, the similar property has been proved in Ref.
\cite{JMP}.

   The transformed Hamiltonian is equal to
$$ {\cal H}'=\rho_3\epsilon+ {\cal E}. $$

   In the general case, the external field is not stationary and the operator
${\cal O}$ commutes neither with ${\cal M}$ nor with ${\cal E}$.
In this case the following transformation method can be used. The
operator ${\cal O}$ can be divided into two operators:
\begin{equation} {\cal O}={\cal O}_1+{\cal O}_2.
\label{eq10} \end{equation}

The operator ${\cal O}_1$ should commute with ${\cal M}$ and the
operator ${\cal O}_2$ should be equal to zero for the free
particle. Therefore, the operator ${\cal O}_2$ should be
relatively small.

First, it is necessary to perform the unitary transformation with
the operator
\begin{equation}
U=\frac{\epsilon+{\cal M}+\rho_3{\cal
O}_1}{\sqrt{2\epsilon(\epsilon+{\cal M})}},~~~
U^{-1}=\frac{\epsilon+{\cal M}-\rho_3{\cal
O}_1}{\sqrt{2\epsilon(\epsilon+{\cal M})}},
~~~\epsilon=\sqrt{{\cal M}^2+{\cal O}^2_1}. \label{eq11}
\end{equation} After this transformation, the Hamiltonian ${\cal H}'$ still contains odd
terms proportional to the derivatives of the potentials. The
operator ${\cal H}'$ can be written in the form \begin{equation}
{\cal H}'=\rho_3\epsilon+{\cal E}'+{\cal O}',~~~\rho_3{\cal
E}'={\cal E}'\rho_3, ~~~\rho_3{\cal O}'=-{\cal O}'\rho_3,
\label{eq12}\end{equation} where $\epsilon$ is defined by Eq.
(11). The odd operator ${\cal O}'$ is small compared to both
$\epsilon$ and the initial Hamiltonian ${\cal H}$. This
circumstance allows us to apply the usual scheme of the
nonrelativistic FW transformation \cite{FW,JMP,BD}.

Second, the transformation should be performed with the following
operator:
\begin{equation} U'=\exp{(iS')}, ~~~
S'=-\frac i4\rho_3\left\{{\cal
O}',\frac{1}{\epsilon}\right\}_+=-\frac i4\left[\frac{\rho_3}{\epsilon},
   {\cal O}'\right]. \label{eq13} \end{equation}
The further
calculations are similar to those performed for spin-1/2 particles
\cite{JMP,BD}. As compared with Ref. \cite{BD}, the particle mass
should be replaced by the operator $\epsilon$ noncommuting with
the operators ${\cal E}'$ and ${\cal O}'$. If only major
corrections are taken into account, then the transformed
Hamiltonian equals
\begin{equation}
{\cal H}''=\rho_3\epsilon+ {\cal E}'+\frac 14\rho_3\left\{\frac{1}
{\epsilon},{{\cal O}'}^2\right\}_+. \label{eq14} \end{equation}

This is the Hamiltonian in the FW representation.

  To obtain the desired accuracy, the calculation
procedure with transformation operator (13) ($S'$ is replaced by
$S'',S'''$, etc.) should be repeated multiply.

It is important that the diagonalization of Hamiltonian is not
equivalent to the FW transformation. There exists an infinite set
of transformations resulting in block-diagonal forms of all the
operators. Therefore, the equivalence of any representation to the
FW one should be verified. For spin-1/2 particles, the example of
the diagonalization which does not lead to the FW representation
has been shown in Ref. \cite{Ydiag}. The similar situation takes
place for spin-1 particles. In particular, the transformation
performed by Roux \cite{Ro} does not lead to the FW representation
either.

\section {Hamiltonian for spin-1 particles in
a nonuniform electromagnetic field}

Young and Bludman \cite{YB} have included terms describing the charge
quadrupole moment of particles in the Corben-Schwinger (CS)
equations \cite{CS} and have made the Sakata-Taketani
transformation \cite{SaTa}. The generalized Sakata-Taketani
equation obtained in Ref. \cite{YB} defines the Hamiltonian acting
on the six-component bispinor. This equation is similar to the
Dirac equation for spin-1/2 particles. Therefore, it is useful to
perform the FW transformation. In this section, we make such a
transformation without an allowance for the charge quadrupole moment
of particles.

The above described method is used for finding the transformed
Hamiltonian to within second-order terms in the field potentials and
first-order terms in the field strengths and first-order
derivatives of the electric field strength. The terms of the
second order and higher in the field strengths and their
derivatives and the first-order terms containing derivatives of
all the orders of the magnetic field strength and derivatives of
the second order and higher of the electric field strength will be
omitted. The external fields are considered to be stationary.

In this approximation, the basic generalized Sakata-Taketani
equation for the Hamiltonian takes the form \cite{YB}
\begin{equation}\begin{array}{c}
{\cal H}=e\Phi+\rho_3 m+i\rho_2\frac{1}{m}(\bm S\cdot\bm
D)^2-(\rho_3+i\rho_2) \frac{1}{2m}(\bm D^2+e\bm S\cdot\bm H)- \\
(\rho_3-i\rho_2) \frac{e\kappa}{2m}(\bm S\cdot\bm H)-
\frac{e\kappa}{2m^2}(1+\rho_1)\biggl[(\bm S\cdot\bm E)(\bm
S\cdot\bm D)-i
\bm S\cdot[\bm E\times\bm D]-\bm E\cdot\bm D\biggr]+ \\
\frac{e\kappa}{2m^2}(1-\rho_1)\biggl[(\bm S\cdot\bm D)(\bm
S\cdot\bm E)-i \bm S\cdot[\bm D\times\bm E]-\bm D\cdot\bm
E\biggr],
\end{array} \label{eq15} \end{equation}
where $\bm H$ is the magnetic field strength, $\kappa=$const, and
$\bm D=\nabla-ie\bm A$.

This equation satisfies Eqs. (7),(10), if
\begin{equation}  \begin{array}{c}
{\cal M}=m+\frac{\bm\pi^2}{2m}-\frac{e}{m}\bm S\cdot\bm H, \\
{\cal E}=e\Phi-\rho_3\frac{e(\kappa-1)}{2m}\bm S\cdot\bm H+\\
\frac{e\kappa}{4m^2}\biggl(\bm S\cdot[\bm\pi\times\bm E]-\bm
S\cdot[\bm E \times\bm\pi]+ \left\{\bm S\cdot\nabla,\bm S\cdot\bm
E\right\}_+-2\nabla\cdot\bm
E\biggr), \\
{\cal O}_1=i\rho_2\left[\frac{\bm\pi^2}{2m}-\frac{(\bm\pi\cdot
\bm S)^2}{m}+\frac{e(\kappa-1)}{2m}\bm S\cdot\bm H\right],\\
{\cal O}_2=i\rho_1\frac{e\kappa}{2m^2}\biggl(\bm\pi\cdot\bm E+\bm
E\cdot\bm\pi- \left\{\bm S\cdot\bm \pi,\bm S\cdot\bm
E\right\}_++\bm S\cdot[\nabla\times\bm E] \biggr),
\end{array} \label{eq16} \end{equation}
where $\bm\pi=-i\bm D=-i\nabla-e\bm A$ is the kinetic momentum
operator and $\left\{\bm S\cdot\nabla,\bm S\cdot\bm
E\right\}_+\equiv (S_iS_j+S_jS_i)(\partial E_i/\partial x_j)$.

The pseudo-unitary FW transformation leads to Eq. (12) where
\begin{equation} \begin{array}{c}
\epsilon=\epsilon'-\left\{\frac{e}{2\epsilon'},\bm S\cdot\bm
H\right\}_+
+\frac{e(\kappa-1)}{16m^2}\left\{\frac{1}{\epsilon'},\{\bm
S\cdot\bm\pi, \bm\pi\cdot\bm H\}_+\right\}_+,  \\
{\cal E}'=e\Phi+\frac{e}{4m}\left[\left\{\left(\frac{\kappa-1}{2}+
\frac{m}{\epsilon'+m}\right)\frac{1}{\epsilon'}, \left(\bm
S\cdot[\bm\pi\times\bm E]-\bm S\cdot[\bm E\times
\bm\pi]\right)\right\}_+- \right. \\ \left. 2\rho_3(\kappa-1)\bm
S\cdot\bm
H-\rho_3\left\{\frac{(\kappa-1)(\epsilon'-m)}{4m\epsilon'(\epsilon'+m)},
\{\bm S\cdot\bm\pi,\bm\pi\cdot\bm H\}_+\right\}_+\right]+\\
\frac{e\kappa}{4m^2}\left\{\left(\bm S\cdot\nabla-
\frac{1}{\epsilon'(\epsilon'+m)}(\bm S\cdot\bm\pi)
(\bm\pi\cdot\nabla)\right),\left(\bm S\cdot\bm E-
\frac{1}{\epsilon'(\epsilon'+m)}(\bm S\cdot\bm\pi)
(\bm\pi\cdot\bm E)\right)\right\}_++\\
\frac{e}{8m^2}\left\{\frac{1}{\epsilon'(\epsilon'+m)}\biggl(\kappa+\frac{m}{\epsilon'+m}\biggr),
\biggl\{\bm S\cdot[\bm\pi\times\nabla],\bm S\cdot[\bm\pi\times\bm E]\biggr\}_+\right\}_+-\\
\frac{e\kappa}{2m^2}\nabla\cdot\bm
E+\frac{e}{4m^2}\left\{\frac{1}{{\epsilon'}^2}
\left(\kappa+\frac{m^2}{4{\epsilon'}^2}\right),
(\bm\pi\cdot\nabla)(\bm\pi\cdot\bm E)\right\}_+, ~~~
\epsilon'=\sqrt{m^2+\bm\pi^2}.
\end{array} \label{eq17} \end{equation}

In this equation the operator ${\cal O}'$ is proportional to the
field strengths. After the second transformation, the contribution
of this operator to the Hamiltonian ${\cal H}''$ is proportional
to ${{\cal O}'}^2$. Such a contribution is negligible and the
Hamiltonian in the FW representation equals
\begin{equation}
{\cal H}''=\rho_3\epsilon+ {\cal E}'
\label{eq18} \end{equation}
or
\begin{equation} \begin{array}{c}
{\cal H}''=\rho_3\epsilon'+e\Phi+\frac{e}{4m}\left[\left\{\left(\frac{\kappa-1}{2}+
\frac{m}{\epsilon'+m}\right)\frac{1}{\epsilon'},
\left(\bm S\cdot[\bm\pi\times\bm E]-\bm S\cdot[\bm E\times
\bm\pi]\right)\right\}_+-
\right. \\ \left.
\rho_3\left\{\left(\kappa-1+\frac{2m}{\epsilon'}\right),
\bm S\cdot\bm H\right\}_++\rho_3\left\{\frac{\kappa-1}{2\epsilon'(\epsilon'+m)},
\{\bm S\cdot\bm\pi,\bm\pi\cdot\bm H\}_+\right\}_+\right]+\\
\frac{e\kappa}{4m^2}\left\{\biggl(\bm S\cdot\nabla-
\frac{1}{\epsilon'(\epsilon'+m)}(\bm S\cdot\bm\pi)
(\bm\pi\cdot\nabla)\biggr),\biggl(\bm S\cdot\bm E-
\frac{1}{\epsilon'(\epsilon'+m)}(\bm S\cdot\bm\pi)
(\bm\pi\cdot\bm E)\biggr)\right\}_++\\
\frac{e}{8m^2}\left\{\frac{1}{\epsilon'(\epsilon'+m)}\left(\kappa+\frac{m}{\epsilon'+m}\right),
\biggl\{\bm S\cdot[\bm\pi\times\nabla],\bm S\cdot[\bm\pi\times\bm E]\biggr\}_+\right\}_+-\\
\frac{e\kappa}{2m^2}\nabla\cdot\bm
E+\frac{e}{4m^2}\left\{\frac{1}{{\epsilon'}^2}
\left(\kappa+\frac{m^2}{4{\epsilon'}^2}\right),
(\bm\pi\cdot\nabla)(\bm\pi\cdot\bm E)\right\}_+.
\end{array} \label{eq19} \end{equation}

     We can introduce the $g$ factor to describe the anomalous magnetic moment (AMM). In this case, $g=\kappa+1$ and
the Hamiltonian takes the form
\begin{equation} \begin{array}{c}
{\cal H}''=\rho_3\epsilon'+e\Phi+\frac{e}{4m}\left[\left\{\left(\frac{g-2}{2}+
\frac{m}{\epsilon'+m}\right)\frac{1}{\epsilon'},
\left(\bm S\cdot[\bm\pi\times\bm E]-\bm S\cdot[\bm E\times
\bm\pi]\right)\right\}_+-
\right. \\ \left.
\rho_3\left\{\left(g-2+\frac{2m}{\epsilon'}\right),
\bm S\cdot\bm H\right\}_++\rho_3\left\{\frac{g-2}{2\epsilon'(\epsilon'+m)},
\{\bm S\cdot\bm\pi,\bm\pi\cdot\bm H\}_+\right\}_+\right]+\\
\frac{e(g-1)}{4m^2}\left\{\biggl(\bm S\cdot\nabla-
\frac{1}{\epsilon'(\epsilon'+m)}(\bm S\cdot\bm\pi)
(\bm\pi\cdot\nabla)\biggr),\biggl(\bm S\cdot\bm E-
\frac{1}{\epsilon'(\epsilon'+m)}(\bm S\cdot\bm\pi)
(\bm\pi\cdot\bm E)\right)\Biggr\}_++\\
\frac{e}{8m^2}\left\{\frac{1}{\epsilon'(\epsilon'+m)}\left(g-1+\frac{m}{\epsilon'+m}\right),
\biggl\{\bm S\cdot[\bm\pi\times\nabla],\bm S\cdot[\bm\pi\times\bm E]\biggr\}_+\right\}_+-\\
\frac{e(g-1)}{2m^2}\nabla\cdot\bm
E+\frac{e}{4m^2}\left\{\frac{1}{{\epsilon'}^2}
\left(g-1+\frac{m^2}{4{\epsilon'}^2}\right),
(\bm\pi\cdot\nabla)(\bm\pi\cdot\bm E)\right\}_+.
\end{array} \label{eq20} \end{equation}

The $g$ factor of $g=g_{Pr}=1$ corresponds to the Proca particle.
Nevertheless, the preferred $g$ factor is equal to 2
\cite{PK,PS,ren}.

   The more particular case
of the uniform electric and magnetic fields has been considered in
Ref. \cite{YP}. Within the nonrelativistic limit, formula (20)
agrees with the result obtained in Ref. \cite{YB}. For the case of
nonrelativistic particles in the magnetic field, this formula
complies with the Hamiltonian derived by Case \cite{C}, who also
used the FW transformation.

Since the electric field is considered to be stationary, $\bm
E=-\nabla\Phi$. Therefore, the operators $\bm S\cdot\nabla$ and
$\bm S\cdot\bm E$ commute. In any case, $\nabla$ operates on $\bm E$.

The transition to the semiclassical approximation consists in
averaging the Hamilton operator over the wave functions of
stationary states. For free particles, the lower spinor is equal
to zero in the FW representation. For particles in external
fields, the maximum ratio of the lower and upper spinors is of the
first order in $W_{int}/E$, where $W_{int}$ is the energy of the
particle interaction with external fields and $E$ is the total energy
of a particle. Thus, we obtain
$(\chi^{\dag}\chi)/(\phi^{\dag}\phi) \sim(W_{int}/E)^2$
\cite{JMP}. Therefore, the contribution of the lower spinor is
negligible and the transition to the semiclassical equation is
performed by averaging the operators in the equation for the upper
spinor. It is usually possible to neglect the commutators between
the coordinate and kinetic momentum operators and between
different components of the kinetic momentum operator (see Ref.
\cite{SJETP2}). As a result, the operator $\bm\pi$ should be
substituted by the classical kinetic momentum. For this classical
quantity we retain the same designation. The semiclassical
Hamiltonian is expressed by the relation
\begin{equation} \begin{array}{c}
{\cal H}''=\epsilon'+e\Phi+\frac{e}{2m}\left[\left(g-2+
\frac{2}{\gamma+1}\right)\biggl(\bm S\cdot[\bm v\times\bm
E]\biggr)- \right. \\
\left.\left(g-2+\frac{2}{\gamma}\right) \bm S\cdot\bm
H+\frac{(g-2)\gamma}{\gamma+1}(\bm S\cdot\bm v)(\bm v\cdot\bm H\}\right]+\\
\frac{e(g-1)}{2m^2}\biggl[\bm S\cdot\nabla-
\frac{\gamma}{\gamma+1}(\bm S\cdot\bm v) (\bm
v\cdot\nabla)\biggr]\biggl[\bm S\cdot\bm E-
\frac{\gamma}{\gamma+1}(\bm S\cdot\bm v)
(\bm v\cdot\bm E)\biggr]+\\
\frac{e\gamma}{2m^2(\gamma+1)}\left(g-1+\frac{1}{\gamma+1}\right)
\biggl(\bm S\cdot[\bm v\times\nabla]\biggr)\biggl(\bm S\cdot[\bm v\times\bm E]\biggr)-\\
\frac{e(g-1)}{2m^2}\nabla\cdot\bm E+\frac{e}{2m^2}
\left(g-1+\frac{1}{4\gamma^2}\right) (\bm v\cdot\nabla)(\bm v
\cdot\bm E),
\end{array} \label{eq21} \end{equation}
where $\gamma=\epsilon'/m$ is the Lorentz factor and $\bm
v=\bm\pi/\epsilon'$ is the velocity. This relation is in the best
compliance with formula (3) defining the Lagrangian for particles
of an arbitrary spin.

Formulae (19)--(21) contain spin-independent terms proportional to
the derivatives of $\bm E$. These terms have not been calculated
in \cite{PK,PS}.

The perfect agreement between Hamiltonian (21) and the Lagrangian
derived in Refs. \cite{PK,PS} causes such an agreement between the
corresponding equations of spin motion. As a result of the
semiclassical transition, the particle polarization operator
reduces to the matrix $\bm S$. The spin motion equation has the
form
\begin{equation}\begin{array}{c}
\frac{d\bm S}{dt}=\left(\frac{d\bm S}{dt}\right)_{BMT}+
\left(\frac{d\bm S}{dt}\right)_{q}, \\
\left(\frac{d\bm S}{dt}\right)_{q}= \frac{Q}{2}\Biggl[\biggl([\bm
S\times\nabla]- \frac{\gamma}{\gamma+1}[\bm S\times\bm v](\bm
v\cdot\nabla)\biggr) \biggl((\bm S\cdot\bm E)-
  \frac{\gamma}{\gamma+1}(\bm S\cdot\bm v)
(\bm v\cdot\bm E)\biggr)+\\ \biggl((\bm S\cdot\nabla)-
\frac{\gamma}{\gamma+1}(\bm S\cdot\bm v)(\bm v\cdot\nabla)\biggr)
\biggl([\bm S\times\bm E]- \frac{\gamma}{\gamma+1}[\bm S\times\bm
v] (\bm v\cdot\bm E)\biggr)\Biggr]+
\\ \frac{e}{2m^2}\frac{\gamma}{\gamma+1}\biggl(g-\frac{\gamma}
{\gamma+1}\biggr)\biggl(\left[\bm S\times [\bm
v\times\nabla]\right] (\bm S\cdot[\bm E\times\bm v])+ \left(\bm
S\cdot [\bm v\times\nabla]\right) \left[\bm S\times[\bm E\times\bm
v] \right]\biggr), \\ Q=-\frac{e(g-1)}{m^2},
\end{array} \label{eq22} \end{equation}
where $\left(\frac{d\bm S}{dt}\right)_{BMT}$ is given by Eq. (2)
and $Q$ is the quadrupole moment.

   Thus, a rigorous calculation shows that, upon the FW transformation,
the Hamiltonian determined on the basis of the Proca theory (with an
allowance for AMM \cite{CS}) is fully consistent with the PKS
theory \cite{PK,PS}. The spin motion equation agrees with the
corresponding equation obtained in Ref. \cite{YP}. Therefore, the
Proca and CS equations correctly describe, at least, weak-field
effects.

\section {Charge quadrupole moment of particles}

Spin-1 particles can possess the charge quadrupole moment. Terms
describing such a moment can be added to the Lagrangian \cite{YB}.
They should be bilinear in the meson field variables $U_\mu$ and
$U_{\mu\nu}$ and linear in the derivatives of the electromagnetic
field $\partial_\lambda F_{\mu\nu}$. The choice of these terms is
strongly restricted by the Maxwell equations. As a result, there
exists the only form of extra terms describing the charge
quadrupole moment of particles \cite{YB}.

The terms that can be added to initial generalized Sakata-Taketani
Hamiltonian (15) are given by
\begin{equation} \begin{array}{c}
\Delta{\cal
H}=\frac{eq}{4m^2}\left[\left(S_iS_j+S_jS_i\right)\frac{\partial
E_i}{\partial x_j}-2\frac{\partial E_i}{\partial x_i}\right]\equiv
\frac{eq}{4m^2}\left[\left\{(\bm S\cdot\nabla),(\bm S\cdot\bm
E)\right\}_+-2\nabla\cdot\bm E\right],
\end{array} \label{eq23} \end{equation}
where $q$=const. These terms should be included into the operator
${\cal E}$. The operator of the unitary transformation defined by
Eq. (11) remains unchanged. As a result of the FW transformation,
Hamilton operator (20) should be added by the terms
\begin{equation} \begin{array}{c}
\Delta{\cal H}''=\frac{eq}{2m^2}\Biggl[(\bm S\cdot\nabla)(\bm
S\cdot\bm E)-\frac{1}{\epsilon'm(\epsilon'+m)^2}(\bm
S\cdot\bm\pi)^2(\bm\pi\cdot\nabla)
(\bm\pi\cdot\bm E)+\\
\frac{\epsilon'-m}{4\epsilon'm(\epsilon'+m)}\left(\biggl\{\bm
S\cdot\bm\pi,(\bm\pi\cdot\nabla)(\bm S\cdot\bm E)\biggr\}_++
\biggl\{\bm S\cdot\bm\pi,(\bm S\cdot\nabla)(\bm\pi\cdot\bm
E)\biggr\}_+ \right)-\nabla\cdot\bm E\Biggr].
\end{array} \label{eq24} \end{equation}

In the nonrelativistic approximation,
\begin{equation} \begin{array}{c}
\Delta{\cal H}''=\frac{eq}{2m^2}\biggl[(\bm S\cdot\nabla)(\bm
S\cdot\bm E)-\nabla\cdot\bm E\biggr].
\end{array} \label{eq25} \end{equation}

This formula is in agreement with the result obtained in Ref.
\cite{YB}. The Hamiltonian describing the quadrupole interaction
of nonrelativistic spin-1 particles is given by \begin{equation}
{\cal H}_q=-\frac{1}{6}Q_{ij}\frac{\partial E_i}{\partial x_j},
~~~ Q_{ij}=\frac{3}{2}Q(S_iS_j+S_jS_i-\frac{4}{3}\delta_{ij}),
\label{eq26}\end{equation} where $Q$ is the quadrupole moment.
With an allowance for Eqs. (22),(25), it is equal to
$$Q=-\frac{e(g-1+q)}{m^2}.$$

The forms of Eqs. (20) and (24) are very different. In the
semiclassical approximation, Hamiltonian (24) is expressed by the
relation
\begin{equation} \begin{array}{c}
\Delta{\cal H}''=-\frac{Q}{2}\Biggl[(\bm S\cdot\nabla)(\bm S\cdot\bm
E)-\frac{\gamma^3}{(\gamma+1)^2}(\bm S\cdot\bm v)^2(\bm
v\cdot\nabla)(\bm v\cdot\bm E)+\\
\frac{\gamma(\gamma-1)}{4(\gamma+1)}\left(\biggl\{\bm S\cdot\bm
v,(\bm v\cdot\nabla)(\bm S\cdot\bm E)\biggr\}_++ \biggl\{\bm
S\cdot\bm v,(\bm S\cdot\nabla)(\bm v\cdot\bm E)\biggr\}_+
\right)-\nabla\cdot\bm E\Biggr],
\end{array} \label{eq27} \end{equation}
where $Q=-eq/m^2$.
Formula (27), unlike formula (21), disagrees with the relativistic
expression for the Lagrangian obtained in Refs. \cite{PK,PS}.

It is evident that formulae (24),(27) does not agree with formulae
(20),(21). The classical description of the quadrupole interaction
of relativistic particles has been given in Ref. \cite{Ycl}. The
results obtained in this work are in agreement with formulae
(20),(21) and contradict formulae (24),(27).

The disagreement between the formulae for the charge quadrupole moment
obtained by different methods poses a difficult problem. Young and
Bludman \cite{YB} used the approach based on an inclusion of
appropriate terms into the first-order Proca Lagrangian. An
addition of first-order terms to the Proca Lagrangian results in the CS
equations which are correct. It is important that there exist the
only form of second-order terms describing the charge quadrupole
moment of particles. However, an inclusion of these terms in the
Lagrangian does not results in a correct relativistic description
of the charge quadrupole moment. As opposed to the Young-Bludman
approach, the PKS one \cite{PK,PS} leads to the correct
second-order Lagrangian. The comparison of this Lagrangian with the classical formulae obtained
in Ref. \cite{Ycl} confirms its validity.

\section {Comparison of classical and quantum formulae for the Hamiltonian}

In the classical theory, the spin motion is usually described by
the well-known Good-Nyborg (GN) equation \cite{G,N}. For the above
used designations, the three-dimensional form of this equation is
given by \cite{G}
\begin{equation} \begin{array}{c}
\frac{d\bm O}{dt}=\left(\frac{d\bm O}{dt}\right)_{BMT}+
\left(\frac{d\bm O}{dt}\right)_{G}, ~~~~~ \bm O=\frac{<\bm S>}{S},\\
\left(\frac{d\bm O}{dt}\right)_{G}= \frac{Q}{2S-1} \biggl((\bm O
\cdot\nabla)+ \frac{\gamma^2}{\gamma+1}(\bm O\cdot\bm v)(\bm
v\cdot\nabla)\biggr) \biggl([\bm O\times\bm E]-  \\
\frac{\gamma}{\gamma+1}[\bm O\times\bm v] (\bm v\cdot\bm E)+\left[
\bm O\times[\bm v\times\bm B]\right]\biggr)+ \\
\frac{egS}{2m^2}\frac{\gamma}{\gamma+1} \biggl[\bm O\times[\bm
v\times\nabla]\biggr] \left[(\bm O\cdot\bm B)-
\frac{\gamma}{\gamma+1}(\bm O\cdot\bm v)(\bm v\cdot \bm B)+
\biggl(\bm O\cdot [\bm E\times\bm v]\biggr)\right],
\end{array} \label{eq28} \end{equation}
where $\left(\frac{d\bm O}{dt}\right)_{BMT}$ is expressed by the
BMT equation.

It is obvious that Eq. (28) disagrees with Eqs. (5) and (22). A
possible reason has been pointed out in Refs. \cite{Pover,Povtw}.
In Refs. \cite{G,N} and some other works the condition of
orthogonality of four-vectors of velocity and polarization ($u^\nu
a_\nu=0$) has been used. However, this condition means the
polarization vector is defined in the particle rest frame
\cite{Pover}. This frame is accelerated. Defining the polarization
vector of particle in the rest frame instead of the instantly
accompanying frame results in changing the spin motion equation
\cite{Pover}. In this case, further calculations are not
well-grounded.

In any case, the use of the Lagrangian or Hamiltonian for defining
the classical equation of spin motion is quite possible. Such an
approach has been used in Refs. \cite{Ycl,Pover,Povtw}. The
Lorentz contraction of longitudinal sizes of moving bodies changes
the Hamiltonian that takes the form
\begin{equation} {\cal H}=\frac16Q_{ij}\frac{\partial^2
\phi}{\partial x_i\partial x_j}+\frac16\tau\frac{\partial^2
\phi}{\partial x_i^2}, \label{eq29} \end{equation} where $Q_{ij}$
is the tensor of quadrupole moment, $\tau$ is the root-mean-square
charge radius, and $x_i~(i=1,2,3)$ are the coordinates of a moving
particle. For nonrelativistic particles
$Q_{ij}=Q_{ij}^{(0)},~\tau=\tau^{(0)}$, where
\begin{equation}
Q_{ij}^{(0)}=\frac{3Q}{2S(2S-1)}\left[S_iS_j+S_jS_i-
\frac23S(S+1)\delta_{ij}\right] \label{eq30} \end{equation} and
$S_i~(i=1,2,3)$ are the spin components. For relativistic
particles \cite{Ycl}
\begin{equation}\begin{array} {l}
Q_{ij}=Q^{(0)}_{ij}-\frac{\gamma}{\gamma+1}\left(v_iv_kQ^{(0)}_{kj}+
v_jv_kQ^{(0)}_{ki}\right)+\frac{\gamma^2}{(\gamma+1)^2}v_iv_jv_k
v_lQ^{(0)}_{lk}+ \\
+\frac{1}{3}\delta_{ij}v_kv_lQ^{(0)}_{lk}-
\left(v_iv_j-\frac{1}{3}\delta_{ij}v^2\right)\tau^{(0)},~~~
\tau=\left(1-\frac{1}{3}v^2\right)\tau^{(0)}-\frac{1}{3}v_kv_lQ^{(0)}_{lk}.
\end{array}
\label{eq31}\end{equation}

The corresponding Hamiltonian takes the form
\begin{equation} \begin{array}{c}
{\cal H}=-\frac{Q}{2S(2S-1)}\biggl[\bm S\cdot\nabla-
\frac{\gamma}{\gamma+1}(\bm S\cdot\bm v) (\bm
v\cdot\nabla)\biggr]\biggl[\bm S\cdot\bm E-
\frac{\gamma}{\gamma+1}(\bm S\cdot\bm v)
(\bm v\cdot\bm E)\biggr]-\\
\frac16\biggl[\tau_0-Q\frac{S+1}{2S-1}\biggr]\biggl[\nabla\cdot\bm
E-(\bm v\cdot\nabla)(\bm v \cdot\bm E)\biggr].
\end{array} \label{eq32} \end{equation}

Classical formula (32) agrees with quantum formulae (3) and (21).
Formulae (3),(21), and (32) give the same relativistic dependence
of terms proportional to the first-order derivatives of the field
strengths. This refers to the terms with and without the spin. The only
difference is an absence of the term proportional to $\biggl(\bm
S\cdot[\bm v\times\nabla]\biggr)\biggl(\bm S\cdot[\bm v\times\bm
E]\biggr)$ in classical Hamiltonian (32). However, this term
describes neither quadrupole interaction nor contact one.
Therefore, it is of a purely quantum origin. Nevertheless, including the
considered term in the classical equation of spin motion is not impossible.

The classical equation of spin motion differing from GN equation
(28) has been derived in Ref. \cite{Ycl}. Use of Hamiltonian (32)
makes it possible to rewrite this equation in the more compact form:
\begin{equation}\begin{array}{c}
\frac{d\bm S}{dt}=\left(\frac{d\bm S}{dt}\right)_{BMT}+
\left(\frac{d\bm S}{dt}\right)_{q}, \\
\left(\frac{d\bm S}{dt}\right)_{q}= \frac{Q}{2}\biggl[\bm
S\times\nabla- \frac{\gamma}{\gamma+1}[\bm S\times\bm v](\bm
v\cdot\nabla)\biggr] \biggl[\bm S\cdot\bm E-
\frac{\gamma}{\gamma+1}(\bm S\cdot\bm v) (\bm v\cdot\bm
E)\biggr]+\\ \frac{Q}{2}\biggl[\bm S\cdot\nabla-
\frac{\gamma}{\gamma+1}(\bm S\cdot\bm v)(\bm v\cdot\nabla)\biggr]
\biggl [\bm S\times\bm E- \frac{\gamma}{\gamma+1}[\bm S\times\bm
v] (\bm v\cdot\bm E)\biggr].
\end{array} \label{eq33} \end{equation}

Eq. (33) agrees with quantum equations (5),(22) and disagrees with
GN equation (28).

Thus, the comparison of the quantum Lagrangian and Hamiltonian
with the classical Hamiltonian derived in Ref. \cite{Ycl} confirms
the validity of both the PKS approach and the CS equations. The GN
equation is not satisfactory.

\section {Wave equations of the second order}

Second-order wave equations can be obtained with an elimination of
several components of wave functions. For example, the second-order
form of Proca equations (1) is the result of substitution of
$U_{\mu\nu}$ into the second equation. After the substitution, the
second-order Proca equation takes the form \cite{Pr}
\begin{equation}
D^\mu D_\mu U_\nu-D^\mu D_\nu U_\mu=m^2 U_\nu. \label{eq34}
\end{equation}

As a rule, the elimination of several components changes
properties of wave functions and wave equations. Wave equations
become non-Hermitian. For example, Eq. (34) is non-Hermitian
because
$$ (D^\mu D_\nu)^\dag=D_\nu D^\mu=D^\mu D_\nu+g^{\mu\rho}[D_\nu,D_\rho]=
D^\mu D_\nu-ieg^{\mu\rho}F_{\rho\nu}\neq D^\mu D_\nu,$$ where
$g^{\mu\rho}={\rm diag}\{1,-1,-1,-1\}$ is the metric tensor and
$F_{\rho\nu}$ is the tensor of electromagnetic field.

A majority of second-order wave equations for spin-1 particles has
been obtained in this way. These equations are non-Hermitian. In
the general case, non-Hermitian equations have complex eigenvalues
and nonorthogonal eigenfunctions. Of course, the elimination of
several components of wave functions changes neither residuary
components nor energy modes. However, only eigenfunctions of
initial Hermitian wave equations are (pseudo)orthogonal. After the
elimination, reduced wave functions become nonorthogonal.
Moreover, the reduction of wave eigenfunctions changes expectation
values of all the operators except for the energy operator. It is
difficult to choose the right set of eigenfunctions because they
are nonorthogonal. Any mistake in choosing eigenfunctions results
in complex energy modes.

In this connection, a presence of complex values in the energy
spectrum of particles in the magnetic field found in Refs.
\cite{TY,GT,T,Ts} is quite natural. In these works, three initial
wave equations have been used. These equations have been
rearranged in appropriate second-order forms. Obtained
second-order wave equations are non-Hermitian. It is no wonder
that corresponding energy modes has been found to be complex.

After the diagonalization, the above equations can be represented
in the following general form:
\begin{equation} \begin{array}{c}
E^2\phi=\left[m^2+\bm\pi^2-e(1+\kappa)\bm S\cdot\bm
H+\frac{e(1-\kappa)} {2m^2}\bm\pi_{\bot}^2(\bm S\cdot\bm
H)+\zeta\right]\phi, \\ \bm\pi_{\bot}= \bm\pi-(\bm\pi\cdot\bm
e_H)\bm e_H,
\end{array} \label{eq35} \end{equation}
where $\zeta$ is the designation for other terms of first and
higher orders in the magnetic field strength and $\bm e_H=\bm
H/H$. The quantities $\zeta$ are different for different equations
(see Refs. \cite{TY,GT,T,Ts}). Eq. (35) can be transformed to the
first-order form by the method proposed in Ref. \cite{PAN} (see
below). In the weak-field approximation,
\begin{equation} \begin{array}{c}
E\phi=\left[\epsilon'-\frac{e(1+\kappa)}{2\epsilon'}\bm S\cdot\bm
H+ \frac{e(1-\kappa)}{4m^2\epsilon'}\bm\pi_{\bot}^2(\bm S\cdot\bm
H)+ \frac{1}{2\epsilon'}\zeta\right]\phi, ~~~
\epsilon'=\sqrt{m^2+\bm\pi^2}.
\end{array} \label{eq36} \end{equation}

First-order equation (36) is consistent neither with the PKS
Lagrangian \cite{PK,PS} nor with the BMT equation \cite{BMT}. It
follows from Eq. (36) that the angular velocity of spin
precession increases when the particle energy increases. This
property also contradicts the BMT equation.

Owing to difficulties discovered in some spin-1 theories, the
problem of their consistency has been posed (see Ref. \cite{VSM}
and references therein). However, the above consideration shows
this problem exists only for second-order spin-1 equations. The
first-order CS equations are fully self-consistent. The problem of
self-consistency of the DKP equation has been investigated in Ref.
\cite{VSM}.

To obtain the correct second-order wave equation, the method
elaborated in Ref. \cite{PAN} can be used. It is based on both the
Feshbach-Villars \cite{FV} and FW transformations. In Ref.
\cite{PAN}, the connection between first-order and second-order
wave equations has been found. The general form of the second-order
wave equation is given by
\begin{equation}
\left[\left(i\frac {\partial}{\partial t}-V\right)^2-(\bm p-\bm
a)^2-m^2\right]\psi=0,\label{eq37} \end{equation}
where the operators $V$ and $\bm a$ characterize the
interaction of a particle with an external
field. These operators can have arbitrary forms and involve the operators of 
coordinate $\bm r$, momentum $\bm p$, and spin $\bm
S$. The wave function $\psi$ is a spinor.

In order to linearize Eq. (37), we introduce the functions $\eta$ and $\zeta$ defined by the
conditions
$$ \psi=\eta+\zeta, ~~~~~ \left(i\frac {\partial}{\partial t}-V\right)\psi=m(\eta-\zeta).
$$

Eq. (37) is equivalent to the following linear equation for the wave function
$\Psi' =\left(\begin{array}{c} \eta \\ \zeta
\end{array}\right)$ (see Ref. \cite{Dav}):
$$i\frac {\partial\Psi'}{\partial t}={\cal H}_0\Psi'=\left[V+\rho_3\left(\frac{{\bm{\pi'}}^2}{2m}+m\right)+ i\rho_2\frac{{\bm{\pi'}}^2}{2m}\right]\Psi',$$
where $\bm{\pi'}=\bm{p}-\bm{a}$ and $\rho_i~(i=1,2,3)$ are the
Pauli matrices.

The Hamiltonian ${\cal H}_0$ can be represented as
\begin{equation}
{\cal H}_0=\rho_3 {\cal M}+V+{\cal O},~~~{\cal M}=\frac{{\bm{\pi'}}^2}{2m}+m, ~~~{\cal O}=i\rho_2\frac{{\bm{\pi'}}^2}{2m},
\label{eq38} \end{equation}
where $V$ and ${\cal O}$ are the even and odd operators,
commuting and anticommuting with $\rho_3$, respectively. This Hamiltonian is pseudo-Hermitian.

In general ($V\neq 0$), Hamiltonian (38) can be transformed to a block-diagonal form in two steps. Assuming that the interaction energy $V$ is small in relation to the total energy of a relativistic particle ($|V|\ll E$), we can first make a transformation with the operator $U$ expressed by Eq. (9). As a result, the Hamiltonian is given by Eq. (12). In this equation, the odd term ${\cal O}'$ is anti-Hermitian.

Since the condition $|{\cal O}'|\ll E$ is now satisfied, we can perform, at the second step, a transformation that is similar to the FW transformation for nonrelativistic particles (see above). If we take into account only the largest corrections, the final Hamiltonian is defined by formula (14).

We retain only terms of order $V^2$ and $\partial^2V/\partial x_i\partial x_j$, disregarding terms proportional to $(\nabla V)^2$ and derivatives of $V$ of the third order and higher. In this approximation, the transformed Hamiltonian takes the form \cite{PAN}
\begin{equation}{\cal H}''=\rho_3\epsilon'+V-
\frac {1}{16}\left\{\frac{1}{{\epsilon'}^4},
(\bm\pi'\cdot\nabla)(\bm\pi'\cdot\nabla)V\right\}_+,\label{eq39} \end{equation} where
$\epsilon'=\sqrt{m^2+{\bm{\pi'}}^2}$.

The inverse problem can be also solved. The first-order wave
equation can be written in the form
\begin{equation}
{\cal H}''=\rho_3\epsilon'+W, \label{eq40}
\end{equation}
where
\begin{equation}
W=V- \frac {1}{16}\left\{\frac{1}{{\epsilon'}^4},
(\bm\pi'\cdot\nabla)(\bm\pi'\cdot\nabla)V\right\}_+.
\label{eq41}\end{equation}

Therefore, the approximate form of the corresponding second-order
wave equation is given by
\begin{equation}
\left[\left(i\frac {\partial}{\partial t}-W- \frac
{1}{16}\left\{\frac{1}{{\epsilon'}^4},
(\bm\pi'\cdot\nabla)(\bm\pi'\cdot\nabla)W\right\}_+\right)^2-
{{\bm\pi}'}^2-m^2\right]\psi=0. \label{eq42} \end{equation}

Use of Eqs. (40)--(42) makes it possible to find the second-order
wave equation for relativistic spin-1 particles interacting with
the electromagnetic field. This equation corresponds to
first-order Eq. (20) and has the form
\begin{equation}\begin{array}{c}
\left[\left(i\frac {\partial}{\partial t}-V\right)^2-
\bm\pi^2-m^2\right]\psi=0, \\
V=e\Phi+\frac{e}{4m}\left[\left\{\left(\frac{g-2}{2}+
\frac{m}{\epsilon'+m}\right)\frac{1}{\epsilon'}, \left(\bm
S\cdot[\bm\pi\times\bm E]-\bm S\cdot[\bm E\times
\bm\pi]\right)\right\}_+- \right. \\ \left.
\rho_3\left\{\left(g-2+\frac{2m}{\epsilon'}\right), \bm S\cdot\bm
H\right\}_++\rho_3\left\{\frac{g-2}{2\epsilon'(\epsilon'+m)},
\{\bm S\cdot\bm\pi,\bm\pi\cdot\bm H\}_+\right\}_+\right]+\\
\frac{e(g-1)}{2m^2}\biggl(\bm S\cdot\nabla-
\frac{1}{\epsilon'(\epsilon'+m)}(\bm S\cdot\bm\pi)
(\bm\pi\cdot\nabla)\biggr)\biggl(\bm S\cdot\bm E-
\frac{1}{\epsilon'(\epsilon'+m)}(\bm S\cdot\bm\pi)
(\bm\pi\cdot\bm E)\biggr)+\\
\frac{e}{4m^2}\left\{\frac{1}{\epsilon'(\epsilon'+m)}\left(g-1+\frac{m}{\epsilon'+m}\right),
\biggl(\bm S\cdot[\bm\pi\times\nabla]\biggr)\biggl(\bm S\cdot[\bm\pi\times\bm E]\biggr)\right\}_+-\\
\frac{e(g-1)}{2m^2}\nabla\cdot\bm
E+\frac{e(g-1)}{4m^2}\left\{\frac{1}{{\epsilon'}^2},
(\bm\pi\cdot\nabla)(\bm\pi\cdot\bm E)\right\}_+, ~~~~~ \bm\pi=\bm
p-e\bm A.
\end{array} \label{eq43} \end{equation}

Eq. (43) is Hermitian. Unfortunately, it is difficult to obtain a
compact four-dimensional form of this equation.

It is evident that Eq. (43) essentially differs from the above
discussed wave equations of the second order. Eq. (43) contains
the three-component wave function $\psi$, whereas wave functions
of other second-order wave equations have four components.

\section {Classification of spin-1 particle interactions}

The results obtained in Refs. \cite{PK,PS,Ycl} and present work
makes it possible to give the complete classification of spin-1
particle interactions. These results are expressed by Eqs. (3),(20),(21), and (32).
First of all, the best agreement between
them can be pointed out. The classical and quantum theories differ only in few
terms of a purely quantum origin.

In Hamilton operators (20) and (21), the first and second terms
are spin-independent. They characterize the interaction of the
charge $e$ with the electromagnetic field.
Lagrangian ${\cal L}_1$ expressed by Eq. (3) and the third terms
in Hamilton operators (20) and (21) describe the electromagnetic
interaction of the magnetic moment of a relativistic particle.

As it known, the preferred value of the factor $g$ is 2 because
only this value makes the quantum electrodynamics be
renormalizable \cite{ren}. The general expression for the magnetic moment is given by
\begin{equation} \mu=\frac{egS}{2m}.\label{eq44} \end{equation} Therefore, the preferred magnetic moment of spin-1 particles equals
\begin{equation} \mu=\mu_0=\frac{e}{m}.
\label{eq45} \end{equation} In the renormalizable electroweak
theory charged vector bosons W$^\pm$ have the magnetic moment
defined by Eq. (45) \cite{ren}.

For spin-1/2 and spin-1 particles, the terms in the Hamiltonians
describing the electromagnetic interactions of the charge and
magnetic moment are rather similar. They differ only due to
different spin matrices.

Spin-1 particles possess the quadrupole moment and
root-mean-square radius. These quantities are nonzero even for
particles that do not have a charge distribution. For
such particles, the operators of quadrupole and contact
interactions are proportional to $g-1$. The electromagnetic
interaction of the quadrupole moment is defined by the first term
in Lagrangian ${\cal L}_2$ and fourth terms in Hamiltonians (20)
and (21). The contact interaction caused by the root-mean-square
radius is given by the sixth and seventh terms in these
Hamiltonians. This interaction has not been considered in Refs.
\cite{PK,PS}. The quadrupole moment and root-mean-square radius
are expressed by the formulae
\begin{equation} Q=-\frac{e(g-1)}{m^2}, ~~~~~ \tau_0=\frac{e(g-1)}{m^2}.\label{eq46} \end{equation}

These formulae has been obtained in Ref. \cite{YB}. Let us note
that the quadrupole operator
$$Q_{ij}=S_iS_j+S_jS_i$$
used by Young and Bludman does not include the contact part $-4\delta_{ij}/3$.

Preferred values of the quadrupole moment and root-mean-square
radius for spin-1 particles are given by the substitution of $g=2$
into Eq. (46). They are equal to
\begin{equation} Q=-\frac{e}{m^2}, ~~~~~ \tau_0=\frac{e}{m^2}.\label{eq47} \end{equation}

These values are attributed to charged vector bosons W$^\pm$. The
preferred value of the quadrupole moment has been obtained in
Refs. \cite{PK,PS,Gian}.

Due to AMMs of particles, the quantities $\mu,~Q$ and $\tau_0$ can be
nonzero even for uncharged particles without any charge
distribution. In this case, it is necessary to replace $eg$ by
$2\mu m$ in all the formulae and then put $e=0$. Extended
particles (charged and uncharged) can also possess the charge
quadrupole moment and charge root-mean-square radius defined by
the charge distribution.

For uncharged spin-1 particles, Eq. (46) takes the form
\begin{equation} Q=-\frac{2\mu}{m}, ~~~~~ \tau_0=\frac{2\mu}{m}.\label{eq48} \end{equation}

Formulae (46)--(48) are valid for particles possessing neither the
charge quadrupole moment nor the charge root-mean-square radius.
For such particles, the relativistic dependence of the quadrupole
and contact interactions is given by formulae (20),(21). As
follows from Eqs. (3) and (32), the relativistic dependence of the
quadrupole interaction remains unchanged for particles with the
charge quadrupole moment. For particles with the charge
root-mean-square radius, an analogous property follows from Eq.
(32).

Eqs. (3),(20),(21) also describe two interactions that do not have
any classical analogues. One of these interactions (the convection
interaction \cite{PK,PS}) is defined by the second term in
Lagrangian ${\cal L}_2$ and the fifth terms in Eqs. (20),(21). The
second interaction is characterized by the term
$$\frac{e}{16}\left\{\frac{1}{{\epsilon'}^4},
(\bm\pi\cdot\nabla)(\bm\pi\cdot\bm E)\right\}_+.$$ This is an
extra term in comparison with the classical expression for the
contact interaction [see formulae (20),(21), and (32)]. This term
has not been calculated in Refs. \cite{PK,PS}. Similar term enters
the Hamiltonian for spin-1/2 particles (the Blount term
\cite{JMP,B}). Both of the interactions vanish in the
nonrelativistic limit.

\section {Discussion and summary}

The above analysis shows the wave equations for spin-1 particles
can be verified. The FW transformation provides a good possibility
of verification. This transformation can be performed for
relativistic particles by the method elaborated in Refs.
\cite{JMP,YP}. The final Hamiltonian is block-diagonal (diagonal
in two spinors).

In the present work, the Hamilton operator in the Foldy-Wouthuysen
representation for relativistic spin-1 particles interacting with
the nonuniform electric and uniform magnetic fields is found. The
more general case of the nonuniform magnetic field is not
considered because of cumbersome calculations. The performed
analysis shows the full agreement between the CS equations
\cite{CS}, the PKS approach \cite{PK,PS}, and the classical theory
\cite{Ycl}. On the contrary, the classical equation of spin motion
found by Good and Nyborg \cite{G,N} is unsatisfactory.

Therefore, the first-order CS equations correctly describe, at
least, weak-field effects. However, the attempt of an allowance
for the charge quadrupole moment fulfilled by adding appropriate
second-order terms to the Lagrangian \cite{YB} does not lead to
the correct result. On the contrary, the PKS approach makes it
possible to find the right Lagrangian for particles of any spin
possessing the charge quadrupole moment. This conclusion poses a
serious problem.

   The Stuckelberg equations also need an analogous investigation. To
find the corresponding spin motion equation, the FW transformation
can be performed for relativistic particles. This transformation
has been made in the nonrelativistic case \cite{V}. Let us mark
that the BMT equation leads to the relation $g_1-g_0=1/2$ between
the coefficients $g_0,g_1,g_2$ used in Ref. \cite{V}.

In contradistinction to Refs. \cite{PK,PS}, Hamiltonians (20),(21)
are calculated with an allowance for spin-independent terms
proportional to $\partial E_i/\partial x_j$, which describe, in
particular, the contact interaction of relativistic particles. The
results obtained in Refs. \cite{PK,PS,Ycl} and the present work
make it possible to give the complete classification of spin-1
particle interactions.

Owing to difficulties discovered in some spin-1 theories, the
problem of their consistency has been posed (see Ref. \cite{VSM}
and references therein). The present work proves this problem exists only for
second-order spin-1 equations. A majority of second-order wave
equations for spin-1 particles has been obtained with an
elimination of several components of wave function. Derived
equations are non-Hermitian. In the general case, non-Hermitian
equations have complex eigenvalues and nonorthogonal
eigenfunctions. The reduction of wave eigenfunctions changes
expectation values of all the operators except for the energy
operator. It is difficult to choose the right set of
eigenfunctions because they are nonorthogonal. Any mistake in
choosing eigenfunctions results in complex energy modes. In this
connection, non-Hermitian wave equations are not quite
satisfactory.

In the present work, the Hermitian second-order wave equation for
spin-1 particles is derived by the method proposed in Ref.
\cite{PAN}. The found equation contains the three-component wave
function, whereas wave functions of other second-order wave
equations have four components.

\section* {Acknowledgements}

I would like to thank Prof. A. V. Borisov for helpful remarks and discussions.

This work was supported by the grant of the Belarusian Republican
Foundation for Fundamental Research No. $\Phi$03-242.

%\footnotesize
%\renewcommand{\baselinestretch}{1.4}


\begin{thebibliography}{}
%\begin{thebibliography}{99}
\bibitem{VSM}
B. Vijayalakshmi, M. Seetharaman and P. M. Mathews, J. Phys. A:
Math. Gen. {\bf 12}, 665 (1979).

\bibitem{PK}
A. A. Pomeransky and I. B. Khriplovich,  JETP {\bf 86}, 839
(1998).

\bibitem{PS}
A. A. Pomeransky and R. A. Sen'kov,  Phys. Lett. B {\bf 468}, 251
(1999).

\bibitem{FW}
L. L. Foldy and S. A. Wouthuysen,  Phys. Rev. {\bf 78}, 29 (1950).

\bibitem{Ycl}
A. J. Silenko,  Yader. Fiz. {\bf 63}, 2139 (2000) [Phys. At.
Nuclei {\bf 63}, 2045 (2000)].

\bibitem{Pr}
A. Proca,  Compt. Rend. {\bf 202}, 1490 (1936).

\bibitem{D}
R. J. Duffin,  Phys. Rev. {\bf 54}, 1114 (1938).

\bibitem{K}
N. Kemmer,  Proc. Roy. Soc. A {\bf 173}, 91 (1939).

\bibitem{Pe}
G. Petiau,  Acad. Roy. Belg. {\bf 16}, 2 (1936).

\bibitem{St}
E. C. G. Stuckelberg,  Helv. Phys. Acta {\bf 11}, 225 (1938).

\bibitem{BW}
V. Bargmann and E. P. Wigner,  Proc. Nat. Acad. Sci. USA {\bf 34},
211 (1946).

\bibitem{CS}
H. C. Corben and J. Schwinger,  Phys. Rev. {\bf 58}, 953 (1940).

\bibitem{YB}
J. A. Young and S. A. Bludman, Phys. Rev. {\bf 131}, 2326 (1963).

\bibitem{Um}
H. Umezawa, {\em Quantum Field Theory} (North-Holland Publ.
Company, Amsterdam, 1956).

\bibitem{Kah}
E. Kahler,  Rendiconti di Matem. {\bf 21}, 425 (1962).

\bibitem{Kr1}
S. I. Kruglov,  Int. J. Theor. Phys. {\bf 41}, 653 (2002); arXiv:
hep-th/0110060.

\bibitem{Kr2}
S. I. Kruglov, {\em Symmetry and Electromagnetic Interaction of
Fields with Multi-Spin} (Nova Sci. Publ., New York, 2001).

\bibitem{Kr3}
S. I. Kruglov, arXiv: hep-th/0304091.

\bibitem{SaTa}
M. Taketani and S. Sakata,  Proc. Phys. Math. Soc. Japan {\bf 22},
757 (1940).

\bibitem{SG}
D. Shay and R. H. Good,  Phys. Rev. {\bf 179}, 1410 (1969).

\bibitem{Ta}
I. E. Tamm,  Doklady Akad. Nauk SSSR (URSS) {\bf 29}, 551 (1940),
{\em in Russian}.

\bibitem{TY}
W. Tsai and A. Yildiz,  Phys. Rev. D {\bf 4}, 3643 (1971).

\bibitem{GT}
T. Goldman and W. Tsai, Phys. Rev. D {\bf 4}, 3648 (1971).

\bibitem{T}
W. Tsai,  Phys. Rev. D {\bf 4}, 3652 (1971).

\bibitem{Ts}
W. Tsai,  Phys. Rev. D {\bf 7}, 1945 (1973).

\bibitem{Zw}
D. Zwanziger,  Phys. Rev. B {\bf 139}, 1318 (1965).

\bibitem{BMT}
V. Bargmann, L. Michel, and V. L. Telegdi,  Phys. Rev. Lett. {\bf
2}, 435 (1959).

\bibitem{YP}
A. J. Silenko,  Yader. Fiz. {\bf 64}, 1054 (2001) [Phys. At.
Nuclei {\bf 64}, 983 (2001)].

\bibitem{CMcK}
J. P. Costella and B. H. J. McKellar,  Am. J. Phys. {\bf 63}, 1119
(1995).

\bibitem{FG}
D. M. Fradkin and R. H. Good,  Rev. Mod. Phys. {\bf 33}, 343
(1961).

\bibitem{JMP}
A. J. Silenko,  J. Math. Phys. {\bf 44}, 2952 (2003).

\bibitem{BD}
J. D. Bjorken and S. D. Drell, {\em Relativistic quantum
mechanics} (McGraw-Hill, New York, 1964).

\bibitem{Ydiag}
A. J. Silenko,  Yader. Fiz. {\bf 61}, 66 (1998) [Phys. At. Nuclei
{\bf 61}, 60 (1998)].

\bibitem{Ro}
J.-F. Roux,  Lett. Nuovo Cim. {\bf 40}, 63 (1984).

\bibitem{ren}
S. Weinberg, in {\em Lectures on Elementary Particles and Quantum
Field Theory}, edited by S. Deser, M. Grisaru, and H. Pendleton
(MIT Press, Cambridge MA, 1970); I. B. Khriplovich,  Zs. Eksp.
Teor. Fiz. {\bf 96}, 385 (1989) [Sov. Phys. JETP {\bf 69}, 217
(1989)]; S. Ferrara, M. Porrati, and V. L. Telegdi, Phys. Rev. D
{\bf 46}, 3529 (1992).

\bibitem{C} K. M. Case,  Phys. Rev.  95, 1323 (1954).

\bibitem{SJETP2}
A. J. Silenko,  Zs. Eksp. Teor. Fiz.  114, 1153 (1998) [JETP {\bf
87}, 629 (1998)].

\bibitem{G}
R. H. Good, Phys. Rev.  125, 2112 (1962).

\bibitem{N}
P. Nyborg, Nuovo Cim. {\bf 31}, 1209 (1964); {\bf 32}, 1131
(1964).

\bibitem{Pover}
A. J. Silenko, Poverchnost', No. 2, 111 (1997) [Surf. Investig.
{\bf 13}, 251 (1998)].

\bibitem{Povtw}
A. J. Silenko, Poverchnost', No. 5, 97 (1998) [Surf. Investig.
{\bf 14}, 687 (1998)].

\bibitem{PAN}
A. J. Silenko, Yader. Fiz. {\bf 64}, 1048 (2001) [Phys. At. Nuclei
{\bf 64}, 977 (2001)].

\bibitem{FV}
H. Feshbach and F. Villars, Rev. Mod. Phys. {\bf 30}, 24 (1958).

\bibitem{Dav}
A. S. Davydov, {\em Quantum
mechanics} (Pergamon, Oxford, 1976).

\bibitem{Gian}
I. Giannakis, J. T. Liu, and M. Porrati, Phys. Rev. D {\bf 58},
045016 (1998).

\bibitem{B} E. I. Blount, Phys. Rev. {\bf 128}, 2454 (1962).

\bibitem{V}
P. A. Vlasov, Ukr. Fiz. Zh. {\bf 27}, 1495 (1977), {\em in
Russian}.

\end{thebibliography}
\end{document}